\documentclass[twocolumn,aps,showpacs,floats,superscriptaddress]{revtex4}
\usepackage[latin9]{inputenc}
\usepackage{amsmath}
\usepackage{amssymb}
\usepackage{graphicx}
\usepackage{esint}

\makeatletter

\providecommand{\tabularnewline}{\\}

\@ifundefined{textcolor}{}
{%
 \definecolor{BLACK}{gray}{0}
 \definecolor{WHITE}{gray}{1}
 \definecolor{RED}{rgb}{1,0,0}
 \definecolor{GREEN}{rgb}{0,1,0}
 \definecolor{BLUE}{rgb}{0,0,1}
 \definecolor{CYAN}{cmyk}{1,0,0,0}
 \definecolor{MAGENTA}{cmyk}{0,1,0,0}
 \definecolor{YELLOW}{cmyk}{0,0,1,0}
}

\usepackage{color}

\newcommand{\be}{\begin{equation}}
\newcommand{\ee}{  \end{equation}}
\newcommand{\ba}{\begin{eqnarray}}
\newcommand{\ea}{  \end{eqnarray}}

\makeatother

\begin{document}

\title{Protected quantum computing: \\
 Interleaving gate operations with dynamical decoupling sequences}

\author{Jingfu Zhang$^{1}$, Alexandre M. Souza$^{2}$, Frederico Dias Brandao$^{1}$,
and Dieter Suter$^{1}$\\
{\it $^1$Fakult$\ddot{a}$t Physik, Technische Universit$\ddot{a}$t
Dortmund, D-44221 Dortmund, Germany}\\
{\it $^2$Centro Brasileiro de Pesquisas Físicas, Rua Dr.Xavier
Sigaud 150, Rio de Janeiro 22290-180, RJ, Brazil}}



\date{\today}
\begin{abstract}
Implementing precise operations on quantum systems is one of the biggest
challenges for building quantum devices in a noisy environment. Dynamical
decoupling (DD) attenuates the destructive effect of the environmental
noise, but so far it has been used primarily in the context of quantum
memories. Here, we present a general scheme for combining DD with
quantum logical gate operations and demonstrate its performance on
the example of an electron spin qubit of a single nitrogen-vacancy
center in diamond. We achieve process fidelities $>$98\% for gate
times that are 2 orders of magnitude longer than the unprotected dephasing
time $T_{2}$.
\end{abstract}


\maketitle
Realizing the potential of quantum computation \cite{nielsen,Stolze:2008xy}
hinges on the implementation of fault-tolerant systems that complete
the computational process with high fidelity even in the presence
of unavoidable environmental perturbations. Quantum error correction
(QEC) offers this possibility, at the cost of an overhead in the number
of qubits, provided that the error per gate can be kept sufficiently
low \cite{zurek98} and the preparation of the initial states is achieved
with sufficiently high fidelity \cite{kitaev2005,laflamme2010}. Achieving
these goals requires additional techniques for eliminating the effect
of perturbations both between and inside the quantum operations. Ideally,
these additional measures should require little or no additional resources.

  Dynamical
decoupling (DD) is an attractive approach for protecting the qubit
system against unwanted environmental interactions, which may be
static or time-dependent. It relies on a sequence of control
operations applied to the system, which refocus the
system-environment interaction. It does not require additional
qubits, and DD sequences can be designed such that they work
reliably even in the presence of unavoidable experimental
imperfections \cite{suter2012,mustafa2013}. Experimental tests of
DD have demonstrated this potential by reducing decoherence rates
in different systems by several orders of magnitude
\cite{bollinger2009,liu2009,suter2010,hanson2010,cory2011,suter2011,suter2011b}.

While most of these tests demonstrated the protection of single qubits
in quantum memories, environmental noise also degrades the fidelity
of quantum gate operations during computational processes\cite{prar2012}.
If the relaxation mechanism is known it is possible to design protected
quantum gates by optimal control techniques \cite{glaser2011}. If
the system environment is not characterized, it may be still possible
to use DD technique for protecting quantum gate operations. In the
simplest case, quantum operation can be made robust against static
noise by refocusing them in a similar manner to a Hahn echo \cite{yacoby2012}.
In the case of a general fluctuating environment, the Hahn Echo must
be replaced by DD methods. Initial experiments demonstrating decoherence
protected quantum gates have been made recently on Nitrogen Vacancy
(NV) Centers \cite{dobrovitski2012} , semiconductor quantum dots
\cite{gossard2010} and solid state nuclear spins \cite{prar2012}.

Possible approaches to build DD protected gates were proposed by several
groups \cite{prar2012,viola2009,viola2010,gyure2010,viola2009b,preskill2011,lukin2009}.
The simplest way to combine DD and gate operations consists of applying
the operations between two consecutive DD cycles. It was theoretically
shown that this approach can lower the resource requirements for QEC
\cite{gyure2010}. However, if the duration of a single gate operation
is comparable to or longer than the decoherence time of the system,
this approach will fail. It becomes then necessary to apply protection
schemes in parallel to the gate operation. This must be done in such
a way that the DD, which is designed to eliminate the effect of interactions
with the environment, does not eliminate the interaction between the
qubits and the control fields driving the gate operation.

In this Letter we demonstrate how it is possible to modify general
logical gate operations in such a way that they can be interleaved
with DD sequences without using auxiliary or encoded qubits. Our method
removes the system-environment interaction for any gate operation
at least to first order and it allows one to combine arbitrary DD
sequences with any type of quantum gate operations.

We consider a system governed by the Hamiltonian
\[
\mathcal{H}(t)=\mathcal{H}_{s}+\mathcal{H}_{c}(t)+\mathcal{H}_{se}+\mathcal{H}_{e},
\]
where $\mathcal{H}_{s}$ describes the internal Hamiltonian of the
qubit, $\mathcal{H}_{c}(t)$ is a time-dependent control
Hamiltonian driving the logical gates, $\mathcal{H}_{se}$ is the
interaction of the qubit with the environment, and
$\mathcal{H}_{e}$ describes the evolution of the environmental
degrees of freedom. Our goal is to implement gate operations
protected against environmental noise. Our target operation is a
unitary gate $U_{t}$ that is not affected by the
system-environment interaction $\mathcal{H}_{se}$:
\[
U_{t}=U_{g}\otimes\mathcal{T}e^{-i\int_{0}^{\tau}\mathcal{H}_{e}dt}.
\]
 Here, the gate operation $U_{g}$ is a pure system operator, $\mathcal{T}$
is the Dyson time ordering operator and $\tau$ is the duration of
the gate operation.

\begin{figure}
\begin{centering}
\includegraphics[width=6cm]{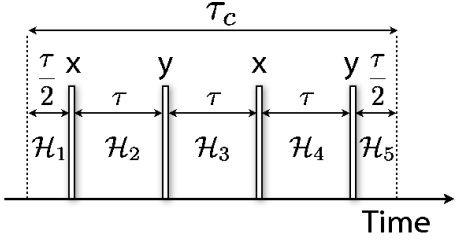}
\par\end{centering}
\caption{Single cycle of an XY-4 DD sequence used as the basis of
protected gate operations. The labels x, y mark the rotation axes
of the DD pulses, and $\mathcal{H}_{n}$ is the Hamiltonian between
the DD pulses.\label{fig:XY4}}
\end{figure}

Protecting the system from the environmental noise while
simultaneously driving logical gate operations can be achieved by
using a standard DD sequence and inserting a suitably adapted gate
operation in short increments in the free precession periods of
the DD sequence. Figure \ref{fig:XY4} illustrates this for the
XY-4 DD sequence: in the free precession periods between the DD
pulses, we insert a control Hamiltonian
$\mathcal{H}_{n}=\mathcal{H}_{s}+\mathcal{H}_{c,n}+\mathcal{H}_{se}+\mathcal{H}_{e}$,
where $(n=1\dots5)$ indicates the period for which this
Hamiltonian is active. The evolution of the system can then be
written as
\begin{equation}
U=U_{N+1}P_{N}U_{N}\dots P_{1}U_{1}=U_{N+1}\Pi_{n=1}^{N}(P_{n}U_{n}),\label{eq:evol1}
\end{equation}
where $N$ is the number of pulses of the DD sequence ($N=4$ in the
case of XY-4), $P_{n}=e^{-i\pi I_{\alpha}}$ is the propagator describing
the $n^{th}$ DD inversion pulse, $I_{\alpha}$ the cartesian component
of the spin operator and $U_{n}=e^{-i\mathcal{H}_{n}\tau_{n}}$ is
the evolution between two DD pulses. We assume that these periods
are short and the control Hamiltonians are time-independent within
each period. We treat the DD pulses $P_{n}$ as ideal rotations.

To find the required control Hamiltonians $\mathcal{H}_{n}$, we re-write
eq. (\ref{eq:evol1}) in the form
\[
U=U_{N+1}\Pi_{n=1}^{N}\tilde{U}_{n}=U_{N+1}\Pi_{n=1}^{N}e^{-i\mathcal{\tilde{H}}_{n}\tau_{_{n}}},
\]
where the Hamiltonians
\[
\tilde{\mathcal{H}}_{n}=T_{n}^{-1}\mathcal{H}_{n}T_{n}
\]
describe the control fields in the so-called toggling frame \cite{haeberlen}
of the DD sequence, which is defined by the transformations
\[
T_{n}=P_{n-1}P_{n-2}\dots P_{1},
\]
which include the limiting cases $T_{1}=T_{N+1}=E$ (identity).
This approach guarantees first order protection to any operation
interlaced with a suitable dynamical decoupling sequence.

\begin{figure}
\centering{}\includegraphics[width=8cm]{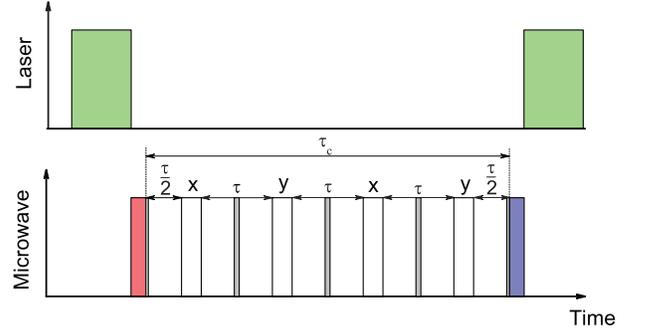}
\caption{Laser (top) and MW (bottom) pulse sequences for the NOT
gate protected by a XY-4 DD cycle. The first laser pulse
initializes the spin into state $|0\rangle$. The second laser
pulse measures the population of state $|0\rangle$. The duration
of the laser pulses is 400 ns. The first MW pulse (red bar)
prepares the input state and the last pulse (blue bar) is the
readout pulse of the quantum process tomography. The pulse
sequence (duration $\tau_{c}$) between the first and last pulse
implements the protected NOT gate, where unfilled boxes represent
the DD pulses, and filled ones the segments of the gate
operation.} \label{fig:wholepulse}
\end{figure}

As a specific example, we choose the XY-4 and XY-8 DD sequences to
protect the gate operations NOOP (no operation, i.e, identity), NOT,
Hadamard and Phase gate, which can be represented as
\[
\left(\begin{array}{cc}
1 & 0\\
0 & 1
\end{array}\right),\quad\left(\begin{array}{cc}
0 & 1\\
1 & 0
\end{array}\right),\quad\frac{1}{\sqrt{2}}\left(\begin{array}{cc}
1 & 1\\
1 & -1
\end{array}\right),\quad\left(\begin{array}{cc}
1 & 0\\
0 & i
\end{array}\right).
\]
To protect these gates, we first split them up into segments that
can be interleaved with the DD sequence. A possible decomposition
is
\begin{eqnarray*}
\mathrm{NOT} & : & (\frac{\pi}{8})_{0}-(\frac{\pi}{4})_{0}-(\frac{\pi}{4})_{0}-(\frac{\pi}{4})_{0}-(\frac{\pi}{8})_{0}\\
\mathrm{H} & : & (\frac{\pi}{4})_{3\pi/2}-(\frac{\pi}{2})_{0}-(0)_{0}-(\frac{\pi}{2})_{0}-(\frac{\pi}{4})_{\pi/2}\\
\mathrm{Phase} & : & (0)_{0}-(\frac{\pi}{2})_{0}-(\frac{\pi}{2})_{\pi/2}-(\frac{\pi}{2})_{0}-(0)_{0}
\end{eqnarray*}
with time running from left to right. Here, $(\theta)_{\phi}=e^{-i\theta(I_{x}\cos\phi-I_{y}\sin\phi)}$
denotes a pulse with flip angle $\theta$ and phase $\phi$. The transformation
into the toggling frame changes the phases to
\begin{eqnarray*}
\mathrm{NOT}_{P} & : & (\frac{\pi}{8})_{0}-(\frac{\pi}{4})_{0}-(\frac{\pi}{4})_{\pi}-(\frac{\pi}{4})_{\pi}-(\frac{\pi}{8})_{0}\\
\mathrm{H}_{P} & : & (\frac{\pi}{4})_{3\pi/2}-(\frac{\pi}{2})_{0}-(0)_{0}-(\frac{\pi}{2})_{\pi}-(\frac{\pi}{4})_{\pi/2}\\
\mathrm{Phase}_{P} & : & (0)_{0}-(\frac{\pi}{2})_{\pi}-(\frac{\pi}{2})_{\pi/2}-(\frac{\pi}{2})_{\pi}-(0)_{0}.
\end{eqnarray*}
The lower part of Fig.\ \ref{fig:wholepulse} shows the resulting
sequence, combining the gate operation and the DD cycle, together
with the red pulse that generates the initial condition, and the readout
pulse (blue).

While we have discussed the example of the XY-4 sequence, the scheme
is equally applicable to other DD schemes. In the case of the XY-8
DD sequence, the protected operation has the general form
\begin{eqnarray*}
 &  & (\theta_{1})_{\phi_{1}}-X-(\theta_{2})_{\phi_{2}}-Y-(\theta_{3})_{\phi_{3}}-\\
 &  & X-(\theta_{4})_{\phi_{4}}-Y-(\theta_{5})_{\phi_{5}}-Y-(\theta_{6})_{\phi_{6}}-X\\
 &  & -(\theta_{7})_{\phi_{7}}-Y-(\theta_{8})_{\phi_{8}}-X-(\theta_{9})_{\phi_{9}},
\end{eqnarray*}
where the flip angles $\theta_{k}$ and phases $\phi_{k}$ are listed
in Table \ref{anglesXY8} and $X=(\pi)_{0}$ and $Y=(\pi)_{\pi/2}$
implement the DD pulses.

\begin{table}
\begin{tabular}{|c|c|c|c|c|c|c|c|c|c|}
\hline
\multicolumn{1}{|c|}{} & $\theta_{1}$  & $\theta_{2}$  & $\theta_{3}$  & $\theta_{4}$  & $\theta_{5}$  & $\theta_{6}$  & $\theta_{7}$  & $\theta_{8}$  & $\theta_{9}$\tabularnewline
\hline
NOT  & $\pi/16$  & $\pi/8$  & $\pi/8$  & $\pi/8$  & $\pi/8$  & $\pi/8$  & $\pi/8$  & $\pi/8$  & $\pi/16$\tabularnewline
\hline
Hadamard  & 0  & $\pi/4$  & $\pi/4$  & $\pi/4$  & 0  & $\pi/4$  & $\pi/4$  & $\pi/4$  & 0\tabularnewline
\hline
Phase  & $0$  & $\pi/4$  & $\pi/4$  & $\pi/4$  & $0$  & $\pi/4$  & $\pi/4$  & $\pi/4$  & $0$\tabularnewline
\hline
\hline
\multicolumn{1}{|c|}{} & $\phi_{1}$  & $\phi_{2}$  & $\phi_{3}$  & $\phi_{4}$  & $\phi_{5}$  & $\phi_{6}$  & $\phi_{7}$  & $\phi_{8}$  & $\phi_{9}$\tabularnewline
\hline
NOT  & 0  & 0  & $\pi$  & $\pi$  & 0  & $\pi$  & $\pi$  & 0  & 0\tabularnewline
\hline
Hadamard  & 0  & $\pi/2$  & $\pi$  & $\pi$  & 0  & $\pi$  & $\pi$  & $3\pi/2$  & 0\tabularnewline
\hline
\multicolumn{1}{|c|}{Phase} & 0  & $\pi$  & 0  & $3\pi/2$  & 0  & $3\pi/2$  & $\pi$  & 0  & 0\tabularnewline
\hline
\end{tabular}\caption{Flip angles ($\theta_{k}$) and phases ($\phi_{k}$) in the gate segments
protected by an XY-8 cycle. }

\label{anglesXY8}
\end{table}

For the experimental test, we used the nitrogen-vacancy (NV) centre
of diamond \cite{6991}, which has an electronic spin $S=1$. Here
we use the subsystem consisting of the $m_{s}=0$ and $+1$ as a single
qubit. We apply a magnetic field along the NV symmetry axis to lift
the degeneracy of the $m_{S}=\pm1$ states. In the secular approximation,
we can write an effective Hamiltonian for the two-level system as
\begin{eqnarray*}
H_{NV} & = & \omega_{S}S_{z}+S_{z}\sum_{j} A_{j}I_{z}^{j}+\sum_{j}\omega_{I}I_{z}^{j}+H_{dip}\\
 & = & \mathcal{H}_{s}+\mathcal{H}_{se}+\mathcal{H}_{e}
\end{eqnarray*}
Here $S_{z}$ and $I_{z}^{j}$ denote the electron and nuclear spin
operators, $\omega_{S}$ and $\omega_{I}$ their resonance frequencies,
$A_{j}$ the hyperfine coupling between the electron and the $j^{th}$
nuclear spin, and $H_{dip}$ the dipolar coupling within the nuclear
spin bath that generates the environmental noise.

In the experiment, we optically address a single NV center using a
green solid-state laser and a home-built confocal microscope. An
acousto-optical modulator with 58 dB extinction ratio and 40 ns
rise-time generates the laser pulses from the CW laser and a 4
GS/s arbitrary waveform generator (AWG) synthesizes the microwave
(MW) pulses at a carrier frequency of 400 MHz. The output of the
AWG is then up-converted by mixing it with the signal from an MW
synthesizer operating at 2.4 GHz. The upper sideband is extracted
by a suitable band pass filter, which attenuates the lower
sideband by 40 dB. The pulses are sent through an 8 W amplifier
and a 20 $\mathrm{\mu m}$ copper wire attached to the diamond
surface.

Figure \ref{fig:wholepulse} illustrates the pulse sequence for implementing
a NOT gate protected by an XY-4 cycle and measuring the performance.
The first laser pulse initializes the spin into state $|0\rangle$.
The second laser pulse implements the measurement of the population
of state $|0\rangle$. The MW pulse sequence is applied between the
two laser pulses. The first MW pulse initializes the state $|0\rangle$
into the input state required for the process tomography, and the
last pulse implements the required readout.

\begin{figure}[h]
\includegraphics[width=8cm]{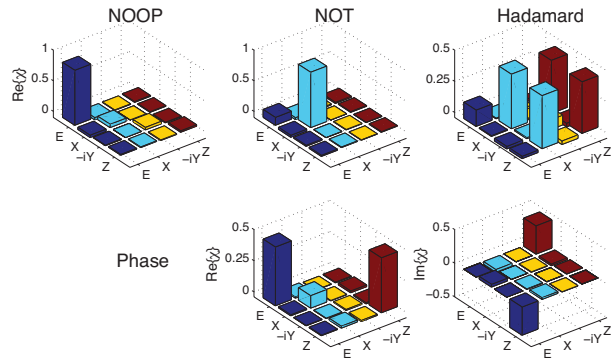} 
\caption{(color online). $\chi$-matrices measured by quantum
process tomography for the NOOP, NOT, Hadamard and Phase gates
protected by XY-8 dynamical decoupling pulses, for gate durations
of $\approx35.5$ $\mu$s. The first row shows the real parts of the
NOOP, NOT and Hadamard gates, and the second row shows real and
imaginary parts of Phase gates, respectively. The imaginary parts
for the NOOP, NOT and Hadamard gates are not shown; their rms
values are $<0.03$, which is compatible with zero within
experimental uncertainties.} \label{Figchi}
\end{figure}

For a quantitative evaluation of the effectiveness of our scheme,
we used quantum process tomography \cite{viola2009} to describe
the process as
$\rho_{out}=\sum_{kl}\chi_{kl}e_{k}\rho_{in}e_{l}^{\dagger}$,
where the basis operators are $e_{k,l}\in\{E,X,iY,Z\}$ and $X$,
$Y$ and $Z$ represent Pauli operators. For each protected gate, we
prepared four states $|0\rangle$, $|1\rangle$,
$(|0\rangle+|1\rangle)/\sqrt{2}$ and
$(|0\rangle-i|1\rangle)/\sqrt{2}$ as the input states. To analyze
the output states, we used quantum state tomography, which
requires four readout operations. Here, we used $E$,
$(\pi/2)_{0}$, $(\pi/2)_{\pi/2}$ and $(\pi)_{0}$. Figure
\ref{Figchi} shows the measured $\chi$ matrices for all four gate
operations protected by the XY-8 sequence, each for a gate
duration of $ \approx35.5$ $\mu$s. For the first three gates,
where the $\chi$-matrices of the ideal gates are real, we only
show the real part. The imaginary parts have rms values of
$0.013$, $0.016$ and $0.027$, respectively.

These matrices prove that the experimentally implemented gates
agree well with the targeted gate operation. We thus conclude that
our method of interleaving gate operations with DD sequences works
and avoids destructive interference between the gate operation and
the DD sequence.

\begin{figure}[h]
\includegraphics[width=8cm]{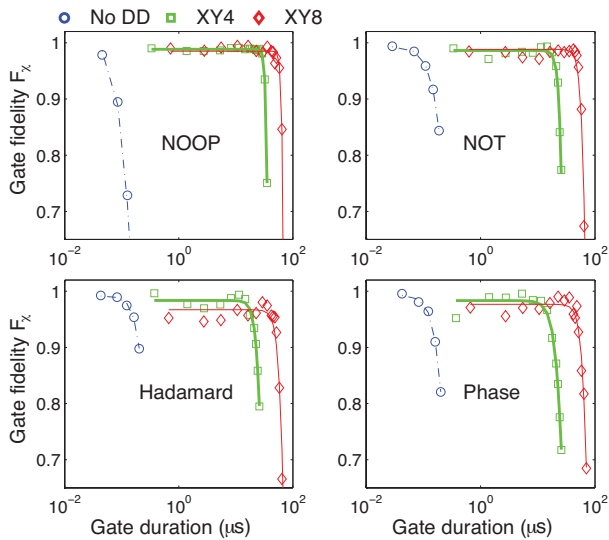} 
\caption{(color online). The measured gate fidelity obtained by quantum process
tomography for the gates without and with dynamical decoupling pulses.
The curves are functions $Ae^{-(t/T_{2})^{k}}$, using the fit parameters
of Table \ref{tab:FitPArs}. }

\label{FigtomoXY4}
\end{figure}

To compare the efficiency of the protection schemes quantitatively,
we determined the gate fidelity from the $\chi$ matrices as \cite{Wang2008}
\begin{equation}
F_{\chi}=|{\rm Tr}(\chi_{exp}\chi_{th}^{\dag})|/\sqrt{{\rm Tr}(\chi_{exp}\chi_{exp}^{\dag}){\rm Tr}(\chi_{th}\chi_{th}^{\dag})}\label{Fchi}
\end{equation}
where $\chi_{th}$ and $\chi_{exp}$ denote the theoretical and experimental
$\chi$ matrices, respectively. For the $\chi$ matrices represented
in Fig.\ \ref{Figchi}, the measured gate fidelities are $0.993$,
$0.985$, $0.975$, $0.989$ for the protected NOOP, NOT, Hadamard
and Phase gates. In figure \ref{FigtomoXY4}, we show how the gate
fidelity changes with increasing gate duration. While the fidelity
of the unprotected gates drops sharply on a timescale of $\approx0.2\,\mathrm{\mu s}$,
the protected gates retain fidelities of the order of $\approx99$
\% for up to 80 $\mathrm{\mu s}$ - clearly demonstrating that the
protection against environmental noise works well also for the gate
operations.

\begin{table}[t]
\centering{}%
\begin{tabular}{|l|c|c|r||||@{\extracolsep{0pt}.}l|r||||@{\extracolsep{0pt}.}l|}
\hline
Gate  & DD cycle  & $A$  & \multicolumn{2}{c|}{$T_{2}$ {[}$\mathrm{\ensuremath{\mu s}}${]}} & \multicolumn{2}{c|}{$k$}\tabularnewline
\hline
NOOP  & -  & 1.01  & \multicolumn{2}{r|}{0.19 } & \multicolumn{2}{r|}{2.6 }\tabularnewline
\hline
 & XY-4  & 0.99  & \multicolumn{2}{r|}{38.2 } & \multicolumn{2}{r|}{17.0 }\tabularnewline
\hline
 & XY-8  & 0.99  & \multicolumn{2}{r|}{71.3 } & \multicolumn{2}{r|}{18.2}\tabularnewline
\hline
NOT  & -  & 0.99  & \multicolumn{2}{r|}{0.36 } & \multicolumn{2}{r|}{2.9}\tabularnewline
\hline
 & XY-4  & 0.99  & \multicolumn{2}{r|}{29.8 } & \multicolumn{2}{r|}{9.6 }\tabularnewline
\hline
 & XY-8  & 0.99  & \multicolumn{2}{r|}{74.7 } & \multicolumn{2}{r|}{7.9 }\tabularnewline
\hline
Hadamard  & -  & 0.99  & \multicolumn{2}{r|}{0.36 } & \multicolumn{2}{r|}{3.9 }\tabularnewline
\hline
 & XY-4  & 0.98  & \multicolumn{2}{r|}{32.4 } & \multicolumn{2}{r|}{6.9 }\tabularnewline
\hline
 & XY-8  & 0.97  & \multicolumn{2}{r|}{77.3 } & \multicolumn{2}{r|}{6.6 }\tabularnewline
\hline
Phase  & -  & 0.99  & \multicolumn{2}{r|}{0.32 } & \multicolumn{2}{r|}{3.5 }\tabularnewline
\hline
 & XY-4  & 0.98  & \multicolumn{2}{r|}{33.2 } & \multicolumn{2}{r|}{4.6 }\tabularnewline
\hline
 & XY-8  & 0.98  & \multicolumn{2}{r|}{83.4 } & \multicolumn{2}{r|}{6.3}\tabularnewline
\hline
\end{tabular}\caption{Summary of experimental gate fidelities for the four gate operations
protected by different DD sequences. The experimental fidelities were
fitted to the function $Ae^{-(t/T_{2})^{k}}$. \label{tab:FitPArs}}
\end{table}

For a quantitative evaluation, we fit the experimental data with the
function $Ae^{-(t/T_{2})^{k}}$ \cite{cory2011,shim12}. Table \ref{tab:FitPArs}
lists the parameters obtained from this fit. Within experimental uncertainty,
the amplitude of all gates is very close to 1.0. The most important
parameter for assessing the effectiveness of the scheme is the decay
time $T_{2}$ of the gate fidelity. Compared to the unprotected gates,
the gates protected by XY-4 extend this lifetime by factors of $201$,
$83$, $89$ and $103$, for NOOP, NOT, Hadamard and Phase gates,
respectively, and the XY-8 scheme achieves factors of $375$, $210$,
$212$ and $258$.

\begin{figure}[h]
\centering{}\includegraphics[width=8cm]{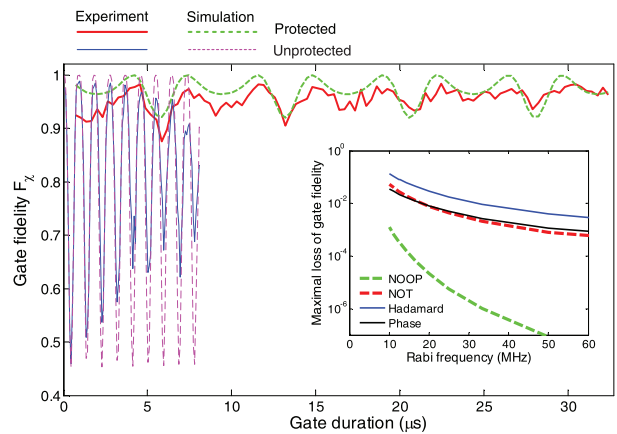} 
\caption{(color online). Effect of the hyperfine coupling between
the electron and the $^{14}$N nuclear spin on the fidelity of the
Hadamard gate. The measured fidelity of the gate protected by an
XY-8 cycle is shown as the full thick curve, and the simulated
fidelity for the same gate as the dashed thick curve. The dashed
thin curve shows the fidelity of the unprotected gate by
simulation and the full thin curve the corresponding experimental
data. The Inset shows how the maximal loss of the gate fidelity
decreases with increasing Rabi frequency of the control pulses.
\label{FigtomoHaddelay}}
\end{figure}

The decay of the gate fidelity in the NV center is dominated by
the hyperfine interaction with the $^{13}$C nuclear spins, which
are present at 1.1 \% of the sites in diamond (natural abundance).
In addition, the electron spin is also coupled to the $^{14}$N
nuclear spin (I=1) of the NV center, through a hyperfine
interaction of $A_{14N}\approx$ $2\pi\cdot2.15\,\mathrm{MHz}$. In
contrast to the nuclear spin bath, this single spin represents a
time-independent perturbation, which also affects the gate
performance, and the coupling strength is significantly larger
than that of the $^{13}$C nuclear spins. In the data shown in
Fig.\ \ref{FigtomoXY4}, we eliminated its effect by an appropriate
choice of the delays between the pulses. In Fig.\
\ref{FigtomoHaddelay}, we explicitly show its effect for the
example of the Hadamard gate. The data shown here correspond to an
expanded scale of the data also represented in the lower left
panel of Fig.\ \ref{FigtomoXY4}, but with higher resolution and
using a linear scale. The oscillations visible in the experimental
as well as the simulated data are due to the hyperfine interaction
between the electronic and the $^{14}$N nuclear spins. The damping
of the experimental oscillations, which is not visible in the
simulated data, can be attributed to the interaction with the
$^{13}$C nuclear spin bath, which was not considered in the
simulations. Clearly, the protection scheme is also helpful for
this type of interaction. In the inset of the figure, we show how
this effect can be eliminated by increasing the Rabi frequency of
the control pulses.

In conclusion, we have introduced a scheme for protecting quantum
logical gate operations against environmental noise by segmenting
the gate operations and interleaving it with a pulse cycle for dynamical
decoupling. The interleaving process requires that the segments of
the gate operations are modified in such a way that the DD pulses
effectively transform them into the operations required by the algorithm.
In the experimental example, using the NV center of diamond, we demonstrated
that protected gates retain high fidelity for durations that are more
than two orders of magnitude longer than for unprotected gates. In
future work, we plan to extend this work to other DD sequences and
to multiqubit systems.

We gratefully acknowledge experimental assistance from J.H. Shim
and useful discussions with G.A. ${\rm\acute{A}}$lvarez. This work
was supported by the DFG through grant Su 192/27-1.


\end{document}